\documentclass[final,5p,times,twocolumn]{ms}

\usepackage{amssymb,amsfonts,amsmath,amstext,amsgen,amsopn,amsxtra,indentfirst,lscape,xtab}
\usepackage{multicol}
\usepackage{blindtext}

\journal{Handbook of Exoplanets}

\begin{document}


\begin{frontmatter}

\title{Radiative Transfer for Exoplanet Atmospheres}

\author[csh]{Kevin Heng}
\ead{kevin.heng@csh.unibe.ch}
\author[nasa]{Mark Marley}
\ead{mark.s.marley@nasa.gov}
\address[csh]{University of Bern, Center for Space and Habitability, Sidlerstrasse 5, CH-3012, Bern, Switzerland}
\address[nasa]{NASA Ames Research Center, Moffett Field, CA 94035, U.S.A.}

\begin{abstract}
Remote sensing of the atmospheres of distant worlds motivates a firm understanding of radiative transfer.  In this review, we provide a pedagogical cookbook that describes  the principal ingredients needed to perform a radiative transfer calculation and predict the spectrum of an exoplanet atmosphere, including solving the radiative transfer equation, calculating opacities (and chemistry), iterating for radiative equilibrium (or not), and adapting the output of the calculations to the astronomical observations.  A review of the state of the art is performed, focusing on selected milestone papers.  Outstanding issues, including the need to understand aerosols or clouds and elucidating the assumptions and caveats behind inversion methods, are discussed.  A checklist is provided to assist  referees/reviewers in their scrutiny of works involving radiative transfer.  A table summarizing the methodology employed by past studies is provided.
\end{abstract}

\end{frontmatter}

\section{What is Radiative Transfer?}

Radiative transfer is the science of how radiation interacts with matter.  In the context of an exoplanet, it describes the passage of radiation that
begins with starlight impinging upon its atmosphere. As the starlight is then either scattered out of the atmosphere or absorbed, we would like to understand how the stellar heating, which depends upon the local balance between the two, is distributed from the top of the atmosphere to the surface of the exoplanet (if it has one).  
We also would like to understand how scattered incident light and thermal emission escapes the atmosphere and travels into the telescope (and detector) of the astronomer.  To build a model of an atmosphere requires that we understand how to propagate radiation between its different layers and how absorbent each layer is.  The ability of each layer to absorb, emit, or scatter light depends on its constituent atoms, molecules and aerosols, which demands that we understand both their chemistry and opacities.

In the current review, our main focus is on understanding the various ingredients that go into performing a radiative transfer calculation, including the assumptions, caveats and techniques.  We discuss how to adapt calculations of a synthetic spectrum to actual data collected by astronomers.  We review key, milestone papers in the context of this discussion and highlight outstanding issues that need to be resolved.  Finally, we provide a checklist for referees/reviewers to better scrutinize papers that involve radiative transfer calculations of exoplanet atmospheres.

\section{A Cookbook for Radiative Transfer: Understanding the Ingredients Involved}

The ingredients in this cookbook are covered in more detail in \cite{seager10}, \cite{marley15}, \cite{hm14} and \cite{heng17}.  Our purpose is to provide a concise overview to aid further reading. As such, we generally aim more to suggest relevant useful papers than to provide definitive citations.

\subsection{Solving the Radiative Transfer Equation}

The study of radiative transfer has its origins in stellar astrophysics.  In essence, the radiative transfer equation is a statement of the conservation of energy \cite{chandra60,mihalas70,mihalas78},
\begin{equation}
\mu \frac{\partial I}{\partial \tau} = I - S.
\end{equation}
This differential equation describes the rate of change of the intensity ($I$), which is the energy per unit area, time, wavelength/frequency/wavenumber and solid angle.  It looks deceivingly simple, because most of the complexity lies hidden in the source function ($S$), which describes the details of absorption, emission, and scattering---both of the external, impinging radiation and thermal emission.

The optical depth ($\tau$) may be visualized as the generalization of length in radiative transfer.  Length alone is not a good indicator of how absorbent an object is, because it also depends on its microphysical properties.  Hidden in the optical depth is the density of atoms, molecules and aerosols, as well as their opacities.  For an atmosphere, we write $\tau = \int \kappa ~d\tilde{m}$ with $\kappa$ being the opacity and $\tilde{m}$ the column mass.

The radiative transfer equation is generally difficult to solve, because $S$ contains an integral that involves what we are solving for in the first place ($I$).  Namely, we have \cite{chandra60,mihalas70,mihalas78}
\begin{equation}
S = \omega \int_{4\pi} I {\cal P} d\Omega^\prime + \left( 1 - \omega \right) B,
\end{equation}
where $\omega$ is the single-scattering albedo (ratio of scattering to total cross sections) and ${\cal P}$ is the scattering phase function, which relates the incident and emergent paths of light.  The integral is performed over all incident angles: $\Omega^\prime$ is the solid angle of incidence.  The Planck function ($B$) accounts for thermal emission due to the medium having a finite temperature.  

The complexity of constructing $S$, so that equation (1) may be solved, lies in the integral over solid angle. To know how much light a finite portion of
the atmosphere may scatter into a component of the radiation field, traveling at a given angle to the local atmosphere, requires knowing the local incident radiation field at all angles, which in turn requires knowing how much light every other layer has already scattered. 
A further complexity is that there are typically two components to the radiation field, which can be visualized by thinking about the daytime sky as seen from the surface of Earth. The incident stellar flux consists of a direct beam (the disk of the sun) as well as a background scattered component (the blue sky,
perhaps punctuated by clouds). Meanwhile, the unseen (for humans) thermal component of the radiation is the much more homogeneously distributed upward and downward fluxes emitted by the ground below and atmosphere above. Treating the radiative transfer, particularly the scattering, of these components with disparate angular distributions is a challenge. 

One solution to the radiative transfer problem is the \textit{Feautrier method}, which combines a pair of radiative transfer equations for the incoming and outgoing intensities into a single, second-order differential equation for the total intensity \cite{mihalas78}.  \cite{hm14} provide a detailed account of this  and other solution methods as well as a discussion of their accuracy and speed. An important
caveat is that the conventional Feautrier method does not permit a general treatment of anisotropic scattering \cite{mihalas78}, but it may be adapted to allow for it (see \cite{sudarsky00} and Section 12.2 of \cite{hm14}).  \cite{sudarsky03} have suggested that the Feautrier method performs poorly for gas giants experiencing intense stellar irradiation.  

In  cases where iterative calculations, and thus numerical efficiency, are required faster radiative transfer methods are desired. A full discussion of
the range of methods is not possible here, but a few popular methods are discussed.

\textit{Two-stream solutions} (e.g., \cite{schuster,mw80}) aim to balance speed, simplicity, and ease of implementation with an acceptance of
lower accuracy.  The key simplification in the two-stream solutions is to solve for the \textit{moments of the intensity}, rather than the intensity itself, which allows us to avoid specifying in detail the functional form of the scattering phase function. 
In general, two-stream solutions are useful for computing angularly-averaged quantities such as heating rates
and albedos. \cite{mw80} provide a comprehensive
summary of various approaches for implementing two-stream approximations.   

Care must be taken that the correct variant of the two-stream solution be applied to a given physical situation. For example, some two-stream methods give unphysical solutions when there is a collimated incident 
beam, such as from the stellar host \cite{mw80}. The delta-discrete ordinates two-stream method \cite{liou80, mw80}, or one of its extensions, is most appropriate in this situation of a discrete incident beam.  This method is used in the irradiated models of \cite{mm99} and  \cite{fortney08}, as well as in subsequent papers by these authors, and it is often employed in the Solar System literature. A recent look at the accuracy of two-stream methods for the calculation
of heating rates in Earth's atmosphere is given in \cite{barker15}. A generalization of the delta-two-stream method, the delta-four-stream method \cite{cuzzi82, liou88}, provides better accuracy in cases of non-isotropic scattering and is used in some general circulation models. Some exoplanet modelers circumvent handling of the direct incident beam by treating the incident flux with a hemispheric average, essentially smearing out the stellar flux
as an upper boundary condition, rather than accounting for the discrete beam, but this is a poor approximation in relatively clear atmospheres and can give spurious heating rates and thermal profiles.

An extension of the two-stream method, appropriate for treating the case of the more uniform thermal emission, is to assume a functional form for the scattering phase function and also assume that the intensity contained in the integral is given by the two-stream solution \cite{toon89}.   This allows for the angular dependence of the 
emergent intensity to still  be computed given the approximation that the initial angular distribution of the  intensity that appears in the scattering source
function is given by the two-stream solution. This  \textit{two-stream source function technique} is well suited to the calculation of layer heating rates
and forms the basis of the thermal component of radiative transfer employed by \cite{marley99}, \cite{fortney08}, \cite{cahoy10} and \cite{morley13}.  Both the two-stream and two-stream source function approaches are able to incorporate non-isotropic or anisotropic scattering \cite{mw80, toon89,hml14}.  Another approach is the \textit{discrete ordinates method}, which approximates the integral in the source function by a series expansion in terms of Legendre polynomials \cite{chandra60}.  This also allows for the treatment of multiple streams but this method can also give unphysical results in some cases \cite{mw80}. \cite{mw80} and \cite{toon89}
present tables showing the accuracy of various two-stream solutions in a variety of both limiting and plausible test cases. 

A number of modern radiative transfer methods have been developed to solve the radiative transfer equation with greater accuracy and flexibility than the two-stream approaches but at the cost of more computational cycles. Several of these are discussed in \cite{hm14}. A method that has its origins in the study of stellar atmospheres is \textit{accelerated lambda iteration}. The method treats the scattering part of the source function iteratively by using appropriate operator-splitting (or preconditioning) methods known from linear algebra \cite{mihalas78,hl95,hm14}.
Iterations continue until any desired level of accuracy is achieved. Chapter 13 of \cite{hm14} provide a detailed historical account and explanation of the accelerated lambda iteration method, as well as figures illustrating the accuracy of the solution as a function of the number of iterations. 

Conceptually, the radiative transfer solution enables us to propagate flux from one model atmospheric layer to the next, treating in detail both the absorption and scattering of external radiation and the thermal emission of the layer.  The choice of radiative transfer solution is often driven by a compromise between accuracy and computational speed. In many practical cases, small errors in the radiation transport are acceptable as they can be dwarfed by other uncertainties in the problem, such as the atmospheric composition, aerosol scattering phase function, opacities, and so on. The relative merits of any particular approach are always a lively topic for discussion by atmospheric modelers.
The appropriateness of any method, however, ultimately rests with the proper implementation for a given problem. One drawback in the modern use of 
standard radiative-transfer routines is that  appreciation of the limits of a given method can be overlooked by casual users. Thus, any user
is advised to investigate the strengths and limitations of any radiative transfer package they may employ.

\subsection{Calculating the Opacities of Atoms and Molecules}

To compute the optical depth and thus perform radiative transfer calculations, we need to know the functional form of $\kappa$.  Gaseous opacities generally depend on temperature, pressure and wavelength \cite{freedman08,freedman14}.  They are essentially a collection of spectral lines, each of which has a strength and a shape.  The strength of each line may be constructed from quantum mechanical quantities such as the Einstein A-coefficient and the oscillator strength \cite{rothman96}.  The classical shapes of spectral lines, accounting for the convolution of the Lorentz and Doppler profiles, are described 
by the Voigt profile (e.g., \cite{draine11}), but the far  
wings of the line profile must account for the pressure broadening from collisions between atoms or molecules (see e.g., Section 3.2 of \cite{hanel92}). Much of the relevant theory
in the astrophysical literature was developed for atoms, but for relatively dense, molecule-rich planetary atmospheres the true line shape remains a complex physics problem \cite{deg14}, which must account for effects such as line mixing. This is a particular issue at very high pressures, where the line shape is difficult to predict accurately. Nevertheless, many hot-Jupiter
exoplanet atmosphere models, particularly for higher gravity planets, extend to depths of 1000 bars or more where the opacity is essentially unknown, an important caveat that is seldom mentioned.

The input data for constructing the spectral lines may be obtained from databases such as \texttt{HITRAN} \cite{rothman96}, \texttt{HITEMP} \cite{rothman10}, \texttt{ExoMol} \cite{tennyson16} or other sources \cite{sb07}. Beyond choosing among sometimes conflicting opacity databases
   and assigning widths for lines in various bands, a particular challenge involved in calculating opacities is computational efficiency \cite{gh15}. 
   This is because one has to compute the line shape for  billions of lines for a set of atoms and molecules across a grid of temperature and pressure---for an integral that is formally indefinite, although a trick exists for evaluating the Voigt profile as a definite integral (e.g., \cite{gh15}).

A true \textit{line-by-line} calculation of radiative transfer would employ opacities that resolved the shape of each and every line.  In practice, we often use a technique borrowed from the atmospheric science community known as the \textit{k-distribution method}, which is capable of reducing an opacity function consisting of billions of lines into a small number of bins ($\sim 10$--1000) each with a smooth, cumulative function \cite{goody89,lo91}.  
The k-distribution method is only exact in the limit of an isothermal, isobaric atmosphere containing one species of atom or molecule.  Otherwise, to use it is to invoke the \textit{correlated-k approximation}; see discussions in \cite{gh15} and \cite{amund17}.  A conceptually simpler approach is to coarsely sample the opacities at discrete wavelengths or frequencies, which we term \textit{opacity sampling}. This approach assumes that if a sufficiently large number of discrete wavelengths are sampled, fluxes can be reliably computed. Different groups rely on various approaches for selecting the sampled wavelengths and the accuracy of the final flux calculation depends on the number and spacing of these wavelengths. 

\subsection{Calculating the Relative Abundances of Atoms and Molecules (Chemistry)}

Carbon, hydrogen, oxygen, nitrogen, sulphur, phosphorus and the other elements are typically sequestered in molecules at the temperatures and pressures encountered in exoplanet atmospheres.  Aerosols may form out of the gas.  Within each layer of the atmosphere, the relative abundances of atoms (e.g., sodium, potassium) and molecules (e.g., water, methane, carbon monoxide) are determined by the temperature, pressure, ultraviolet irradiation environment (for photochemistry) and atmospheric mixing (which delivers material from other layers).  It is helpful to visualize the system as a network of reactants and products, some of which are transient.  Each of the links between nodes in this network may be described by a chemical timescale.  When a specific molecule (e.g., methane) is converted to another molecule (e.g., carbon monoxide), one may compute an effective chemical timescale through this network associated with the conversion.  When this effective timescale is shorter than the dynamical mixing timescales of the atmosphere, then the layer is in chemical equilibrium.

Chemical equilibrium may be computed using a procedure known as \textit{Gibbs free energy minimization} \cite{zeggeren}, which is the chemical analog of minimizing the Lagrangian in Newtonian mechanics.  Only the reactants and products matter, and not the pathways taken (e..g, \cite{bs99,lf02,madhu12,hlt16}), which renders computation comparatively easy.  However, if the chemical and dynamical timescales are comparable, then one needs to solve a large set of mass conservation equations with source and sink terms, analogous to solving the equations of motion in classical mechanics (e.g., \cite{moses11,tsai17}).  This is known as \textit{chemical kinetics}. In some cases, including the canonical methane-to-carbon-monoxide equilibration, the precise intermediate molecular pathways are not perfectly known, so conversion timescales can be inherently uncertain (see the discussion in \cite{zahn14}).

\subsection{Cloud and Haze Opacity}

Every Solar System planet with a substantial atmosphere is home to clouds and hazes, generally termed aerosols. Exoplanet atmospheres are no exception (e.g., \cite{heng16}). Properly handling the radiative transfer in the presence of atmospheric aerosols adds another layer of complexity to the calculation. A common
approach is to treat aerosols with Mie theory, which implements the exact solution of Maxwell's equations in idealized spheres. If the composition is known, the real and imaginary indices of refraction can be employed to compute the scattering and absorption efficiencies of spheres of a given size. Monodisperse spheres, however, exhibit strong interference effects in the efficiencies as a function of wavelength, but such effects are seldom seen in nature as they are 
washed out in the presence of even a modest distribution of particle sizes or realistic irregular shapes. Thus, it is important to include a range of particle sizes---and never only a single
size---in almost all realistic cases \cite{hansen74}. A variety of different size distributions are commonly used, including log-normal, power-law, Hansen-Hovenier, and others. Further complexities, such as non-spherical or layered particles, can be introduced but such refinements are unlikely to be warranted by the available exoplanet datasets in the near future.

With Mie theory, one can compute the angular dependence of scattered radiation and this could rigorously be used in the radiative transfer solution, but it is instead common practice to approximate the exact calculation with an analytical approximation, such as the Henyey-Greenstein phase function, for computational speed---often in conjunction with a two-stream calculation. However, this approximation can also introduce errors of 5--10\% in computed quantities; see the appendix of \cite{barker15} for an example.

The difficulty in any forward modeling application is determining the location, number densities, sizes, and optical properties of any 
atmospheric aerosols. The properties of hazes from photochemical processes and clouds from the condensation of atmospheric species can
be estimated of course, but the presence of these opacity sources alters the radiation budget of an atmosphere, requiring further iteration to find
a self-consistent solution. A more detailed review of such issues can be found in \cite{marley13}.

\subsection{Iterating for Radiative Equilibrium (Or Not)}

Opacities depend on temperature, but the temperature structure of an atmosphere also depends on the opacities.  Clearly, some form of iteration has to occur to obtain a converged model.  How is this iteration performed?  What is the condition of convergence?  Physically, from conservation of energy
we expect that at \textit{radiative equilibrium} the sum of all radiative fluxes into and out of an atmospheric layer to be equal to zero. The sum 
must account for energy absorbed and emitted by the layer, as well as the passage of radiation through the layer from the incident flux and any internal heat flow---plus the incident flux absorbed below the layer. This means that if a given layer absorbs incident flux, either from stellar radiation or from the thermal flux emitted from below, the layer must heat up until it can radiate the amount of energy it absorbs while transmitting the unabsorbed flux from below. If a layer absorbs incident flux but the gasses in that layer are poor radiators in the thermal infrared, the layer temperature must increase
until the layer can emit adequate flux to balance that absorbed.  For a simple parameterization, see \cite{robinson14}. 

To find the atmospheric structure that satisfies these requirements requires multiple iterations of the temperature profile.
This iteration continues until the temperature reaches a steady state and radiative equilibrium is established.  Radiative equilibrium is a statement of the \textit{local}, rather than the global, conservation of energy.  Local energy conservation implies global energy conservation, but not vice versa \cite{hl16}.  Whether radiative equilibrium is an appropriate condition to enforce in model atmospheres remains an active topic of debate. Departures from
radiative equilibrium can occur when other energy transport mechanisms, such as convection or breaking of atmospheric gravity waves, are important. When convection is important, the atmospheric thermal profile must be iteratively adjusted as well, e.g., see discussion in \cite{marley15}.

\subsection{Adapting Your Calculations to Astronomical Observations}

A self-consistent, converged radiative transfer calculation yields the radiation escaping from the top of the model atmosphere---the predicted spectrum, which is the flux as a function of wavelength or frequency.  Such predictions may be directly compared to spectra of exoplanet atmospheres. The exact nature of the comparison depends upon the type of exoplanet data that is being modeled.

 For transiting exoplanets, the appropriate comparison depends on whether one is interested in transmission or emission spectra.  For emission spectra, it is the ratio of the flux of the exoplanet atmosphere to that of the star that is measured.  It requires knowledge of the spectrum of the star.  Typically, the \texttt{Kurucz} \cite{kurucz79} or \texttt{PHOENIX} \cite{ah95} stellar models are used.  For transmission spectra, it is the ratio of the radius of the exoplanet to that of its star that is measured.  The transit radius corresponds to the chord cutting across the exoplanet atmosphere that has an optical depth on the order of unity \cite{fortney05}.  The transmission spectrum is calculated by locating the transit radius at each wavelength \cite{brown01,h01}.  Transmission spectra are relatively insensitive to temperature, although the strength of spectral features in transmission are sensitive to pressure scale height, which depends on temperature, as well as gravity and mean molecular mass.

For directly imaged exoplanets, the quantity of interest is often ``contrast" or the ratio of the wavelength-dependent brightness to that of the star, which must be known. Because high-contrast coronagraphic observations discard the light of the star itself, the planet flux must be inferred from the contrast ratio and
the separately observed or modeled stellar flux. In this sense, direct-imaging observations are not unlike thermal-emission observations of transiting planets.

\section{Highlights of the State of the Art}

\begin{figure}
\begin{center}
\vspace{-0.1in}
\includegraphics[width=0.9\columnwidth]{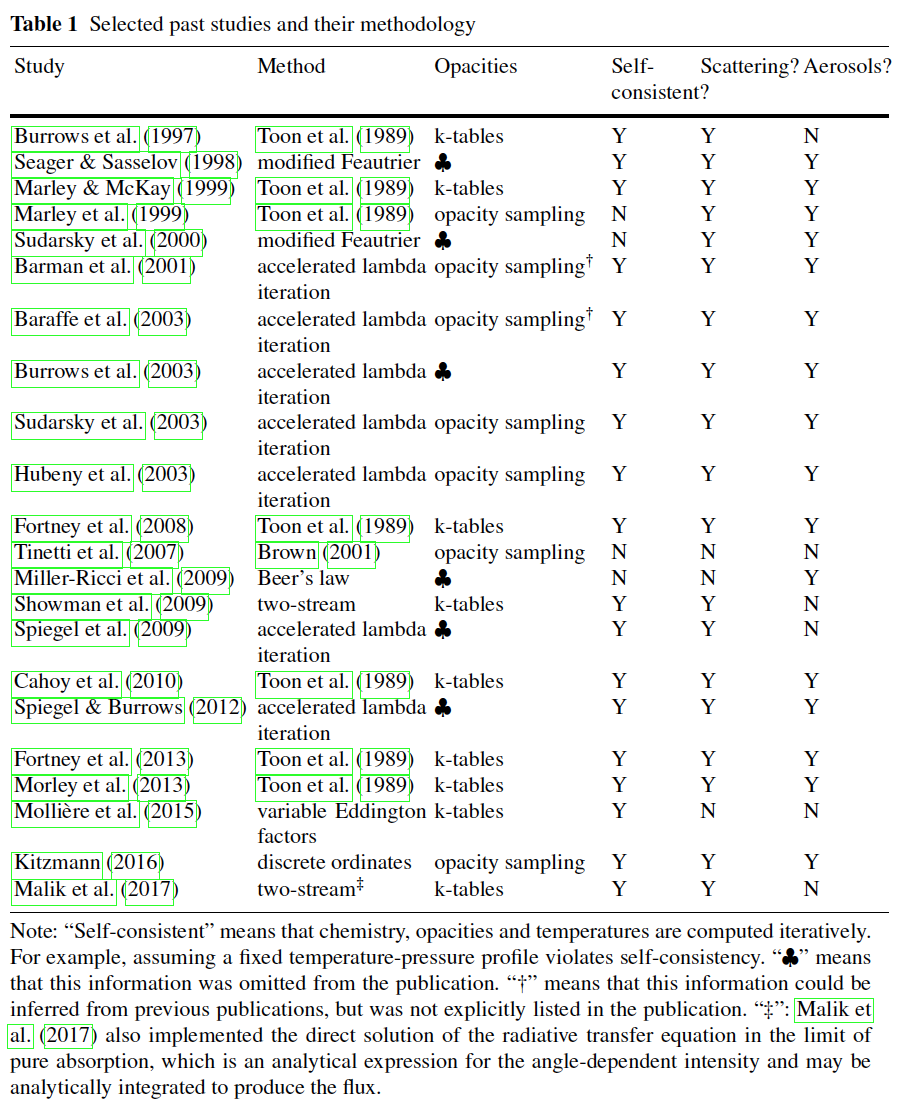}
\end{center}
\vspace{-0.3in}
\label{tab:studies}
\end{figure}

We focus on the papers that laid the foundation for the application of radiative transfer in subsequent papers concerning exoplanet atmospheres, while recognizing that there is a substantial literature on stars, substellar objects and brown dwarfs (which we do not review here).

\subsection{Milestones}

Broadly speaking, the radiative transfer codes being used in the literature can be traced back to having a planetary science or brown dwarf (and stellar) origin.  \cite{ss98} were the first to calculate self-consistent models of irradiated gas giants using a Feautrier-method code originally developed for binary stars, demonstrating that intense stellar irradiation renders the photospheric regions more isothermal.

At about the same time, \cite{marley98} published thermal-structure models of a few extrasolar giant planets using a giant-planet-atmosphere model originally developed to study the atmosphere of Uranus \cite{mm99}. This treatment of incident solar and emitted thermal fluxes
was based on the radiative transfer scheme of  \cite{toon89} as originally applied to Titan by \cite{mckay89}, and
formed the foundation for later studies by this group (e.g., \cite{fortney05}). Using this approach, \cite{fortney08} supported
the prediction of \cite{hub03} that two classes of hot Jupiters exist based on the absence or presence of temperature inversions in their atmospheres as mediated by titanium and vanadium oxide, although today there is only tentative evidence for the detection of titanium or vanadium oxide in one hot Jupiter, WASP-121b \cite{evans16}.
Other early examples of planetary-science-heritage radiative transfer being applied to exoplanet science are the work of \cite{gouk00} and \cite{Iro05}.


Early studies of brown dwarf atmospheres, which led directly to subsequent exoplanet models, included the work of: \cite{marley96} and \cite{burrows97}, which built on the planetary \cite{mm99} formalism; \cite{allard96}, which grew from the stellar atmospheres studies of \cite{ah95}. \cite{burrows97} coupled the radiative transfer calculations  to 
interior models to compute evolutionary tracks for brown dwarfs and young gas-giant planets.
Later, \cite{burrows03} adapted the \texttt{TLUSTY} computer code, which has a heritage from stellar atmospheres \cite{Hubeny88, hl95}, towards studying hot Jupiters.  \cite{sudarsky03} studied gas giants at different orbital distances from their stars and elucidated the influence of stellar irradiation on their spectra.  \cite{burrows07} furthered the debate on temperature inversions in hot Jupiters and the identity of the unknown species causing these inversions.  \cite{spiegel09} used the same computational setup to suggest that titanium oxide may cause temperature inversions only if the atmospheric mixing is strong enough to replenish it against rainout.  \cite{sb12} modeled gas giants with different initial entropies (cold, warm and hot starts) and made predictions for how these scenarios may be distinguished observationally.

\cite{sudarsky00} coined the term ``roasters" for hot Jupiters and predicted low albedos in the range $900 \le T_{\rm eq} \le 1500$ K (where $T_{\rm eq}$ is the equilibrium temperature), but increase when $T_{\rm eq} \ge 1500$ K.  The reasoning is that the condensation curves of silicates (e.g., enstatite) intersect the temperature-pressure profiles of these $T_{\rm eq} \ge 1500$ K atmospheres at higher altitudes and conceivably produce aerosol layers that have high albedos.  Observations of the geometric albedo have instead indicated that it is somewhat constant with equilibrium temperature, suggesting that the formation of aerosol layers is a more complicated process \cite{hd13}.

\cite{barman01} used the \texttt{PHOENIX} code to compute grids of model atmospheres for irradiated gas giants and concluded that the presence of aerosols affects both the temperature structure and emergent spectrum.  \cite{baraffe03} suggested that the inflated radius of HD 209458b cannot be caused by stellar irradiation alone, and that an extra source of energy is required for its inflated radius.

\cite{showman09} recognized that the tidally-locked, highly-irradiated hot Jupiters required the coupling of radiative transfer to three-dimensional fluid dynamics.  They coupled a general circulation model (GCM) to the radiative transfer code of \cite{mm99}, but employed the two-stream approximation and k-distribution tables to boost computational efficiency.  \cite{showman09} was the first study to confront synthetic spectra and phase curves from a GCM with measured data.

Special attention has also been given to the calculation of transmission spectra \cite{brown01,h01,fortney05}.  \cite{ss00} predicted that the sodium and potassium lines would be readily detectable, a prediction that has been verified by subsequent observations (e.g., \cite{sing16}).  \cite{burrows03} emphasized the importance of distinguishing between the photospheric and transit radii\footnote{Order-of-magnitude considerations already inform us that the pressure level corresponding to the transit radius should be $\sim \sqrt{H/R}$ smaller than the photospheric pressure, where $H$ is the pressure scale height and $R$ is the transit radius.}.  \cite{ricci09} elucidated the strength of features in the transmission spectra of super Earths and its relationship to the mean molecular mass of the atmosphere. \cite{fortney13} suggested that the transmission spectra of low-density super Earths and mini-Neptunes could be flat because of high mean molecular weight or photochemical hazes, and proposed observational tests for how to distinguish between these two scenarios. Recently, \cite{robinson17} investigated the importance multiple scattering, which had previously been neglected, may play in certain transit-spectroscopy observations.

\subsection{Outstanding Issues}

The traditional modeling approach is to combine all of the ingredients we described, which involves making a set of assumptions, and performing radiative transfer calculations to make predictions on the temperature structure and spectrum of an exoplanet atmosphere.  An alternative approach, which has its origins in the study of the Earth's atmosphere, is to solve the inverse problem instead: starting from the measured transmission or emission spectrum, can we back out the chemistry and temperature structure \cite{ms09}?  This approach is known as \textit{atmospheric retrieval}.  Generally, one needs to make a compromise between the accuracy of the \textit{representation} of physics and chemistry versus the need for computational efficiency.  The parameter space of a model exoplanet atmosphere is potentially vast and multi-dimensional, and one of the challenges involved with retrieval is to thoroughly explore this parameter space, not get trapped in local minima, and understand the biases and degeneracies involved.  Ideally, one would like to back out the physics and chemistry given the data.  In reality, there are systematic biases implicitly built into the model used, because of the simplifying assumptions applied---one is instead backing out the \textit{model} given the data.

One of the challenges related to radiative transfer is to understand, from first principles, how aerosols form and evolve, and to include their radiative influence into one's calculation.  In principle, one starts from a set of elemental abundances, predicts the composition of the gas, condenses out seed particles from the gas and evolves them to form a size distribution of aerosols---probably of heterogeneous composition---and accounts for their radiative effects on the atmosphere self-consistently \cite{hell06}. 
This is a problem at the interface of microphysics, atmospheric modeling, and radiative transfer. Whether approaching the problem from
a forward modeling or retrieval perspective, some simplifications must be made or else one runs the risk of drowning in free parameters. Identifying
the right simplifications that will inform, but not overwhelm, is a forefront problem in exoplanet atmosphere modeling.

\section{A Checklist for Referees and/or Careful Readers}

When reviewing a paper, either as a referee or a reader, it is often difficult to obtain complete knowledge of the radiative transfer method used because of omitted information.  To address this issue, we introduce the following checklist of questions.
\begin{itemize}

\item What is the form of the radiative transfer equation being solved?  Is the geometry plane-parallel or spherical?  Are there simplifying approximations being invoked (e.g., pure absorption limit)?

\item What is the method of solution of the radiative transfer equation (e.g., two-stream, Feautrier)?

\item How is the incident stellar flux assumed to be treated at the top of the model, i.e., as a discrete beam or a smooth distribution in angle?

\item For the opacities, what are the atoms and molecules being considered?  What are the spectroscopic databases being used to construct the opacities?  Is collision-induced absorption included?  Is pressure broadening included, and if so are the spectral line wings being truncated at a certain wavenumber?  Has a treatment for the continuum opacity been applied to compensate for the truncation of these line wings? How are widths assigned to individual lines within various bands and are they appropriate for $\rm H_2$ (or whichever gas is dominant in the atmosphere) broadening? 

\item What is the method used to include the opacities in radiative transfer?  E.g., k-distribution method with correlated-k approximation, opacity sampling.

\item What is the spectral resolution being used to sample the opacities?  Have the authors explicitly performed convergence tests to demonstrate that the resolution is sufficient?  Numerical convergence and spectral resolution of the measured data are separate issues.  A model should first demonstrate convergence, and then be binned down to the spectral resolution of the data.   

\item If the authors claim to perform a ``line-by-line" treatment of radiative transfer, have they explicitly stated their spectral resolution of the opacities?  Does this spectral resolution qualify as being line-by-line?  As an example, one naively expects that a Gaussian curve cannot be resolved with one sampling point.

\item Is scattering being included?  Is the scattering isotropic or anisotropic?  Has a specific functional form been assumed for the scattering phase function?

\item Is an aerosol model included?  It is possible to include scattering without including aerosols.  It is also possible to partially include the effects of aerosols by specifying a finite Bond albedo, but this does not account for the fact that scattering will alter the \textit{shape} of the temperature-pressure profile \cite{hhps12}.

\item Is a chemistry model included?  Has chemical equilibrium been assumed?  Is photochemistry included?  If a chemical kinetics code is being used (to treat disequilibrium chemistry), what is the value or vertical profile being assumed for the eddy mixing/diffusion coefficient?

\item For chemical kinetics, what is the size of the chemical network being employed?  How many species and chemical reactions have been considered?  Has it been demonstrated that the computer code reproduces chemical equilibrium as a sanity check?

\item For photochemistry, what is the range of ultraviolet wavelengths being considered?  What is the spectral resolution of this wavelength grid?  Have the authors demonstrated that their results are robust to the spectral resolution of their ultraviolet wavelength grid?

\end{itemize}

As a resounding example of how the radiative transfer method used may affect the conclusion drawn, \cite{kitzmann16} revisited the atmospheric modeling of early Mars using a discrete ordinates method and found that two-stream calculations over-estimated the \textit{scattering} greenhouse warming by several tens of degrees, because the two-stream treatment is inaccurate when large\footnote{``Large" has a well-defined meaning in radiative transfer: it means that the size of the aerosol is much larger than the wavelength of light it is scattering.} aerosols are involved.  In the case of Mars, the large aerosols take the form of carbon dioxide ice clouds.  The more accurate calculation using the discrete ordinates method produced atmospheric models with temperatures below the freezing point of water, suggesting that carbon dioxide ice clouds alone are insufficient for maintaining liquid water on the surface of early Mars.

As we transition into a higher-precision era of comparative exoplanetology, a natural next step for the exoplanet atmospheres community to take is to elucidate the assumptions, caveats and limitations of the theoretical tools used and how they influence the conclusions drawn.

\section{Acknowledgements}

The authors thank J. Cuzzi, I. Hubeny, J. Fortney, D. Kitzmann, C. Morley, T. Robinson and R. Freedman for helpful comments on drafts of this entry.  KH acknowledges partial financial support from the PlanetS National Center of Competence in Research (NCCR), the Swiss National Science Foundation, the Center for Space and Habitability and the Swiss-based MERAC Foundation.



\begin{thebibliography}{99}

\bibitem[Allard \& Hauschildt(1995)]{ah95} Allard, F., \& Hauschildt, P. H. \ 1995, Astrophysical Journal, 445, 433

\bibitem[Allard et al.(1996)]{allard96} Allard, F., Hauschildt, P.~H., Baraffe, I., \& Chabrier, G.\ 1996, Astrophysical Journal Letters, 465, L123 

\bibitem[Amundsen et al.(2017)]{amund17} Amundsen,D., et al. \ 2017, Astronomy \& Astrophysics, 598, A97

\bibitem[Barman et al.(2001)]{barman01} Barman, T.S., Hauschildt, P.H., \& Allard, F. \ 2001, Astrophysical Journal, 556, 885

\bibitem[Baraffe et al.(2003)]{baraffe03} Baraffe, I., et al. \ 2003, Astronomy \& Astrophysics, 402, 701

\bibitem[Barker et al.(2015)]{barker15} Barker, H.~W., Cole, J.~N.~S., Li, J., Yi, B., \& Yang, P.\ 2015, Journal of Atmospheric Sciences, 72, 4053 

\bibitem[Brown(2001)]{brown01} Brown, T.M. \ 2001, Astrophysical Journal, 553, 1006

\bibitem[Burrows et al.(1997)]{burrows97} Burrows, A., et al. \ 1997, Astrophysical Journal, 491, 856

\bibitem[Burrows \& Sharp(1999)]{bs99} Burrows, A., \& Sharp, C.M. \ 1999, Astrophysical Journal, 512, 843

\bibitem[Burrows et al.(2003)]{burrows03} Burrows, A., Sudarsky, D., \& Hubbard, W.B. \ 2003, Astrophysical Journal, 594, 545

\bibitem[Burrows et al.(2007)]{burrows07} Burrows, A., et al. \ 2007, Astrophysical Journal Letters, 668, L171

\bibitem[Cahoy et al.(2010)]{cahoy10} Cahoy, K.L., Marley, M.S., \& Fortney, J.J. \ 2010, Astrophysical Journal, 724, 189

\bibitem[Chandrasekhar(1960)]{chandra60} Chandrasekhar, S. \ 1960, Radiative Transfer (New York: Dover)

\bibitem[Cuzzi et al.(1982)]{cuzzi82} Cuzzi, J.N., Ackerman, T.P., \& Helmle, L.C. \ 1982, Journal of Atmospheric Sciences, 39, 917

\bibitem[Draine(2011)]{draine11} Draine, B.T. \ 2011, Physics of the Interstellar and Intergalactic Medium (New Jersey: Princeton University Press)

\bibitem[Evans et al.(2016)]{evans16} Evans, T.M., et al. \ 2016, Astrophysical Journal Letters, 822, L4

\bibitem[Fortney(2005)]{fortney05} Fortney, J.J. \ 2005, Monthly Notices of the Royal Astronomical Society, 364, 649

\bibitem[Fortney et al.(2008)]{fortney08} Fortney, J.J., et al. \ 2008, Astrophysical Journal, 678, 1419

\bibitem[Fortney et al.(2013)]{fortney13} Fortney, J.J., et al. \ 2013, Astrophysical Journal, 775, 80

\bibitem[Freedman et al.(2008)]{freedman08} Freedman, R.S., Marley, M.S., \& Lodders, K. \ 2008, Astrophysical Journal Supplements, 174, 504

\bibitem[Freedman et al.(2014)]{freedman14} Freedman, R.S., et al. \ 2014, Astrophysical Journal Supplements, 214, 25

\bibitem[Goody et al.(1989)]{goody89} Goody, R., West, R., Chen, L., \& Crisp, D.\ 1989, Journal of Quantitative Spectroscopy and Radiative Transfer, 43, 191 

\bibitem[Grimm \& Heng(2015)]{gh15} Grimm, S.L. \& Heng, K. \ 2015, Astrophysical Journal, 808, 182

\bibitem[Goukenleuque et al.(2000)]{gouk00} Goukenleuque, C., B{\'e}zard, B., Joguet, B., Lellouch, E., \& Freedman, R.\ 2000, Icarus, 143, 308 

\bibitem[Hanel et al.(1992)]{hanel92} Hanel, R.~A., Conrath, B.~J., Jennings, D.~E., \& Samuelson, R.~E.\ 1992, Exploration of the solar system by infrared remote sensing, Cambridge Planetary Science Series, No. 7 (Cambridge: Cambridge University Press)

\bibitem[Hansen \& Travis(1974)]{hansen74} Hansen, J.~E., \& Travis, L.~D.\ 1974, Space Science Reviews, 16, 527 

\bibitem[Helling \& Woitke(2006)]{hell06} Helling, Ch., \& Woitke, P. \ 2006, Astronomy \& Astrophysics, 455, 325

\bibitem[Heng et al.(2012)]{hhps12} Heng, K., et al. \ 2012, Monthly Notices of the Royal Astronomical Society, 420, 20

\bibitem[Heng \& Demory(2013)]{hd13} Heng, K., \& Demory, B.-O. \ 2013, Astrophysical Journal, 777, 100

\bibitem[Heng et al.(2014)]{hml14} Heng, K., Mendon\c{c}a, J.M., \& Lee, J.-M. \ 2014, Astrophysical Journal Supplements, 215, 4

\bibitem[Heng et al.(2016)]{hlt16} Heng, K., Lyons, J.R., \& Tsai, S.-M. \ 2016, Astrophysical Journal, 816, 96

\bibitem[Heng \& Lyons(2016)]{hl16} Heng, K., \& Lyons, J.R. \ 2016, Astrophysical Journal, 817, 149

\bibitem[Heng(2016)]{heng16} Heng, K. \ 2016, Astrophysical Journal Letters, 826, L16

\bibitem[Heng(2017)]{heng17} Heng, K. \ 2017, Exoplanetary Atmospheres: Theoretical Concepts and Foundations (New Jersey: Princeton University Press)

\bibitem[Hubbard et al.(2001)]{h01} Hubbard, W.B., Fortney, J.J., Lunine, J.I., Burrows, A., Sudarsky, D., \& Pinto, P. \ 2001, Astrophysical Journal, 560, 413

\bibitem[Hubeny(1988)]{Hubeny88} Hubeny, I.\ 1988, Computer Physics Communications, 52, 103

\bibitem[Hubeny \& Lanz(1995)]{hl95} Hubeny, I., \& Lanz, T. \ 1995, Astrophysical Journal, 439, 875

\bibitem[Hubeny \& Mihalas(2015)]{hm14} Hubeny, I., \& Mihalas, D. \ 2015, Theory of Stellar Atmospheres (New Jersey: Princeton University Press)

\bibitem[Hubeny et al.(2003)]{hub03} Hubeny, I., Burrows, A., \& Sudarsky, D.\ 2003,  Astrophysical Journal, 594, 1011 

\bibitem[Iro et al.(2005)]{Iro05} Iro, N., B{\'e}zard, B., \& Guillot, T.\ 2005, Astronomy \& Astrophysics, 436, 719 

\bibitem[Kitzmann(2016)]{kitzmann16} Kitzmann, D. \ 2016, Astrophysical Journal Letters, 817, L18

\bibitem[Kurucz(1979)]{kurucz79} Kurucz, R.L. \ 1979, Astrophysical Journal Supplements, 40, 1

\bibitem[Lacis \& Oinas(1991)]{lo91} Lacis, A.A., \& Oinas, V. \ 1991, Journal of Geophysical Research, 96, 9027

\bibitem[Liou(1980)]{liou80} Liou, K.N., An Introduction to Atmospheric Radiation (New York: Academic Press)

\bibitem[Liou(1988)]{liou88} Liou, K.-N., Fu, Q., \& Ackerman, T.P. \ 1988, Journal of Atmospheric Sciences, 45, 1940

\bibitem[Lodders \& Fegley(2002)]{lf02} Lodders, K., \& Fegley, B. \ 2002, Icarus, 155, 393

\bibitem[McKay et al.(1989)]{mckay89} McKay, C.~P., Pollack, J.~B., \& Courtin, R.\ 1989, Icarus, 80, 23 

\bibitem[Madhusudhan \& Seager(2009)]{ms09} Madhusudhan, N., \& Seager, S. \ 2009, Astrophysical Journal, 707, 24

\bibitem[Madhusudhan(2012)]{madhu12} Madhusudhan, N. \ 2012, Astrophysical Journal, 758, 36

\bibitem[Malik et al.(2017)]{malik17} Malik, M., et al. \ 2017, Astronomical Journal, 153, 56

\bibitem[Marley et al.(1996)]{marley96} Marley, M.S., et al.\ 1996, Science, 272, 1919 

\bibitem[Marley(1998)]{marley98} Marley, M.S.\ 1998, Brown Dwarfs and Extrasolar Planets (ASP Conference Series), 134, 383 

\bibitem[Marley \& McKay(1999)]{mm99} Marley, M.S., \& McKay, C.P. \ 1999, Icarus, 138, 268

\bibitem[Marley \& Robinson(2015)]{marley15} Marley, M.S., \& Robinson, T.~D.\ 2015, Annual Review of Astronomy \& Astrophysics, 53, 279 

\bibitem[Marley et al.(1999)]{marley99} Marley, M.S., et al. \ 1999, Astrophysical Journal, 513, 879

\bibitem[Marley et al.(2013)]{marley13} Marley, M.S., Ackerman, A.~S., Cuzzi, J.~N., \& Kitzmann, D.\ 2013, Comparative Climatology of Terrestrial Planets, 610, 367 (Tucson: University of Arizona Press)

\bibitem[Meador \& Weaver(1980)]{mw80} Meador, W.E., \& Weaver, W.R. \ 1980, Journal of the Atmospheric Sciences, 37, 630

\bibitem[Mihalas(1970)]{mihalas70} Mihalas, D. \ 1970, Stellar Atmospheres, first edition (San Francisco: Freeman and Company)

\bibitem[Mihalas(1978)]{mihalas78} Mihalas, D. \ 1978, Stellar Atmospheres, second edition (San Francisco: Freeman and Company)

\bibitem[Miller-Ricci et al.(2009)]{ricci09} Miller-Ricci, E., Seager, S., \& Sasselov, D. \ 2009, Astrophysical Journal, 690, 1056

\bibitem[Molli\`{e}re et al.(2015)]{mol15} Molli\`{e}re, P., et al. \ 2015, Astrophysical Journal 813, 47

\bibitem[Morley et al.(2013)]{morley13} Morley, C.V., et al. \ 2013, Astrophysical Journal, 775, 33

\bibitem[Moses et al.(2011)]{moses11} Moses, J.I., et al. \ 2011, Astrophysical Journal, 737, 15

\bibitem[Robinson \& Catling(2014)]{robinson14} Robinson, T.~D., \& Catling, D.~C.\ 2014, Nature Geoscience, 7, 12 

\bibitem[Robinson(2017)]{robinson17} Robinson, T.D. \ 2017, Astrophysical Journal, 836, 236

\bibitem[Rothman et al.(1996)]{rothman96} Rothman, L.S., et al. \ 1996, Journal of Quantitative Spectroscopy \& Radiative Transfer, 60, 665

\bibitem[Rothman et al.(2010)]{rothman10} Rothman, L.S., et al. \ 2010, Journal of Quantitative Spectroscopy \& Radiative Transfer, 111, 2139

\bibitem[Schuster(1905)]{schuster} Schuster, A. \ 1905, Astrophysical Journal, 21, 1

\bibitem[Seager \& Sasselov(1998)]{ss98} Seager, S., \& Sasselov, D.D. \ 1998, Astrophysical Journal Letters, 502, L157

\bibitem[Seager \& Sasselov(2000)]{ss00} Seager, S., \& Sasselov, D.D. \ 2000, Astrophysical Journal, 537, 916

\bibitem[Seager(2010)]{seager10} Seager, S. \ 2010, Exoplanet Atmospheres: Physical Processes (New Jersey: Princeton University Press)

\bibitem[Sharp \& Burrows(2007)]{sb07} Sharp, C.M., \& Burrows, A. \ 2007, Astrophysical Journal Supplements, 168, 140

\bibitem[Showman et al.(2009)]{showman09} Showman, A.P., et al. \ 2009, Astrophysical Journal, 699, 564

\bibitem[Sing et al.(2016)]{sing16} Sing, D.K., et al. \ 2016, Nature, 529, 59

\bibitem[Spiegel et al.(2009)]{spiegel09} Spiegel, D.S., Silverio, K., \& Burrows, A. \ 2009, Astrophysical Journal, 699, 1487

\bibitem[Spiegel \& Burrows(2012)]{sb12} Spiegel, D.S., \& Burrows, A. \ 2012, Astrophysical Journal, 745, 174

\bibitem[Sudarsky et al.(2000)]{sudarsky00} Sudarsky, D., Burrows, A., \& Pinto, P. \ 2000, Astrophysical Journal, 538, 885

\bibitem[Sudarsky et al.(2003)]{sudarsky03} Sudarsky, D., Burrows, A., \& Hubeny, I. \ 2003, Astrophysical Journal, 588, 1121

\bibitem[Tennyson et al.(2014)]{deg14}Tennyson, J., et al. \ 2014, Pure and Applied Chemistry, 86, 1931 (arXiv:1409.7782)

\bibitem[Tennyson et al.(2016)]{tennyson16} Tennyson, J., et al. \ 2016, Journal of Molecular Spectroscopy, 327, 73

\bibitem[Tinetti et al.(2007)]{tinetti07} Tinetti, G., et al. \ 2007, Astrophysical Journal Letters, 654, L99

\bibitem[Toon et al.(1989)]{toon89} Toon, O.B., McKay, C.P., \& Ackerman, T.P. \ 1989, Journal of Geophysical Research, 94, 16287

\bibitem[Tsai et al.(2017)]{tsai17} Tsai, S.-M., et al. \ 2017, Astrophysical Journal Supplements, 228, 20

\bibitem[van Zeggeren(1970)]{zeggeren} van Zeggeren, F., \& Storey, S.H. \ 1970, The Computation of Chemical Equilibria (New York: Cambridge University Press)

\bibitem[Zahnle \& Marley(2014)]{zahn14} Zahnle, K.~J., \& Marley, M.~S.\ 2014, Astrophysical Journal, 797, 41 

\end{thebibliography}
\end{document}